\documentclass[graybox]{svmult}

\usepackage{natbib}

\usepackage{booktabs}
\usepackage{type1cm}        
\usepackage{makeidx}         
\usepackage{graphicx}        
\usepackage{multicol}        
\usepackage[bottom]{footmisc}

\usepackage{newtxtext}       %
\usepackage{newtxmath}       

\usepackage{hyperref}

\makeindex             

\newcommand{\MyPara}[1]{\vspace{.2em}\noindent\textit{\textbf{#1}}\hspace{.3em}}

\begin{document}

\title*{Rapid Reviews in Software Engineering}
\author{Bruno Cartaxo, Gustavo Pinto, and Sergio Soares}
\institute{Bruno Cartaxo \at Federal Institute of Pernambuco (IFPE), \\Paulista, Pernambuco, Brazil,
\\\email{email@brunocartaxo.com}
\and Gustavo Pinto \at Federal University of Par\'a (UFPA),
\\Bel\'em, Par\'a, Brazil
\\\email{gpinto@ufpa.br}
\and Sergio Soares \at Federal University of Pernambuco (UFPE),
\\Recife, Pernambuco, Brazil
\\\email{scbs@cin.ufpe.br}}

\maketitle


\abstract{Integrating research evidence into practice is one of the main goals of Evidence-Based Software Engineering (EBSE). Secondary studies, one of the main EBSE products, are intended to summarize the best research evidence and make them easily consumable by practitioners. However, recent studies show that some secondary studies lack connections with software engineering practice. In this chapter, we present the concept of Rapid Reviews, which are lightweight secondary studies focused on delivering evidence to practitioners in a timely manner. Rapid reviews support practitioners in their decision-making, and should be conducted bounded to a practical problem, inserted into a practical context. Thus, Rapid Reviews can be easily integrated in a knowledge/technology transfer initiative. After describing the basic concepts, we present the results and experiences of conducting two Rapid Reviews. We also provide guidelines to help researchers and practitioners who want to conduct Rapid Reviews, and we finally discuss topics that my concern the research community about the feasibility of Rapid Reviews as an Evidence-Based method. In conclusion, we believe Rapid Reviews might interest researchers and practitioners working in the intersection between software engineering research and practice.}

\section{Introduction}
\label{sec:introduction}

Evidence-Based Practice aims to curate the best research evidence in a given domain of expertise and integrate the findings into practice~\citep{mckibbon1998evidence}. The medical research field was one of the pioneers embracing such a paradigm. More recently, following the promising results in medicine, many other research fields have been adopting Evidence-Based Practice, such as: psychology~\citep{Anderson2006}, nursing~\citep{Dicenso1998}, crime prevention~\citep{Farrington2003}, social work~\citep{Webb2001}, and education~\citep{davies1999evidence}. The seminal paper of ~\cite{Kitchenham2004} introduced the Evidence-Based Practice in the software engineering community. According to the authors, the goal of the Evidence-Based Software Engineering (EBSE) is:

\begin{quote}
\textit{``to provide the means by which current best evidence from research can be \textbf{integrated with practical experience} and human values in the decision-making process regarding the development and maintenance of software.''}~\citep{Kitchenham2004} (bold emphasis added)
\end{quote}

Considering this goal, it is no coincidence that secondary studies 
are the main product of EBSE. Some authors argue that the knowledge aggregated in secondary studies is the most appropriate to be transferred to practice~\citep{Lavis2003}. This belief is rooted in years of Evidence-Based Practice, showing that individual studies often lead to different conclusions compared to more mature and comprehensive secondary studies~\citep{Lavis2003}. As an example, a study comparing the mortality rates of for-profit and nonprofit hospitals found a lower risk of death in for-profit hospitals. On the opposite direction, a secondary study, considering data from studies that summed up 26,000 hospitals and 38 millions of patients, found a higher risk of death in for-profit hospitals~\citep{Devereaux2002}.

Fast forwarding 15 years, EBSE is now a mature field with new studies being conducted on regular basis~\citep{daSilva2011,Borges2014,Borges2015}. However, despite its evolution, several researchers claim that EBSE still lacks connection with software engineering practice~\citep{Hassler2014,Santos2013,daSilva2011}. An investigation with researchers specialized in EBSE revealed that the ``lack of connection with industry'' is the sixth top barrier to conduct secondary studies, from a total of 37 barriers~\citep{Hassler2014}. In the same direction, the study of ~\cite{Santos2013} deployed a survey to 44 authors of 120 secondary studies; only six of them affirmed their studies had direct impact on industrial practice. In addition, a tertiary study identified that only 32 out of 120 secondary studies provide guidelines to practitioners. These findings may indicate that EBSE has not been accomplishing its main goal.

The Evidence-Based Medicine community also faced similar problems in its early days, and it is still facing to some extent nowadays~\citep{Best1997,tricco2017rapid,Tricco2015}. To mitigate this lack of connection with practice, one of the most successful initiatives of the medical field is what has been called Rapid Reviews (RRs)~\citep{Tricco2015}. They are secondary studies aiming to provide research evidence to support decision-making in practice. RRs must be conducted taking into account the constraints inherent to practical environments, such as time and effort. RRs usually deliver evidence in a more timely manner, with lower costs, and reporting results through more appealing mediums~\citep{Cartaxo2018ease}. As a consequence, RRs tend to be more connected to practice, when compared to Systematic Reviews (SRs)\footnote{By SRs we mean the more methodologically rigorous secondary studies, like: meta-analyzes, the traditional systematic literature reviews, and systematic mapping studies~\citep{Kitchenham2007}}. To achieve these goals, RRs omit or simplify some steps of SRs. For instance, RRs can limit the search sources or use just one person to screen primary studies~\citep{Tricco2015}.

Inspired by our peers from the medical field, we recently employed the concept of RRs in software engineering contexts~\citep{Cartaxo2018ist,Cartaxo2018ease,cartaxo2019esem}. The kick start of a RR is a practical problem that exist in a software project. This particular problem must motivate researchers to screen the literature looking for potential answers. As a consequence, researchers must work closely to practitioners to guarantee that the RR is close tied to a practical context. Instead of using a traditional paper-based format, the results of a RR should be incorporated in more attractive mediums, such as Evidence Briefings, which are one-page documents reporting the main findings of a RR~\citep{Cartaxo2016briefings}. 


At first sight, one may argue that while RRs speed up the process by simplifying some predefined steps of SRs, it may also introduce methodological threats. To better understand this concern, some studies have been conducted in medicine to evaluate the impact of RRs methodological adaptations, in comparison to SRs~\citep{abou2016methods,corabian2002rapid,Best1997,Taylor-Phillips2018,Van2011}. Although there are evidence reporting divergences between RRs and SRs~\citep{Van2011}, there are more evidence reporting the similarity of results obtained with those two approaches \citep{abou2016methods,corabian2002rapid,Best1997,Taylor-Phillips2018}. While further investigations are still needed to draw more conclusive results, RRs should not be understood as a replacement to SRs. Instead, we believe that both can (and should) co-exist: while SRs are important to provide in-depth evidence, RRs are useful to easily and quickly transfer scientific knowledge to practice.

In this chapter, we introduce the background concepts related to RRs (Sect.~\ref{sec:what_is_rr}); show results and experiences on conducting such kind of studies in software engineering (Sect. \ref{sec:examples}); introduce guidelines on how to plan, perform and report RRs (Sect. \ref{sec:process}); present further discussions about topics that may concern software engineering research community about the feasibility of RRs (Sect. \ref{sec:further_discussions}); list recommended further reading (Sect. \ref{sec:further_reading}); and present the conclusions (Sect. \ref{sec:conclusion});

\section{Background}
\label{sec:what_is_rr}

In this section there are some background information about what is a RR; why using RRs, based on evidence of their benefits; who is using RRs; and how RRs compares to SRs in terms of their results and methodological characteristics.

\subsection{What is a Rapid Review?}
\label{subsec:what_is_rr}

Rapid Reviews are practice-oriented secondary studies~\citep{watt2008,Haby2016,polisena2015rapid,tricco2017rapid}. The main goal of a RR is to provide evidence to support decision-making towards the solution, or at least attenuation, of issues practitioners face in practice. To support this goal and to meet practice time constraints, RRs have to deliver evidence in shorter time frames, when compared to SRs, which often take months to years~\citep{Tricco2015}. To make RRs compliant with such characteristics, some steps of SRs are deliberately omitted or simplified.

Since RRs are a recent phenomenon in Evidence-Based Medicine, many methodological variations have been identified. This can be observed in the study of~\cite{featherstone2015advancing}, which analyzed the methods employed in many published RRs. Additionally, \cite{tricco2016international} interviewed 40 RRs producers and also observed the presence of method variability. These two studies identified high heterogeneity among RRs, from varying time frames, to ambiguous definitions of what is a RR. Despite RRs high methodological variability, the majority RRs share at least the following core aspects:

\vspace{0.2cm}
\begin{shaded}
\MyPara{Rapid Reviews should be performed in close collaboration with practitioners, bounded to practical problems, and conducted within practitioners context:} The argument to conduct lightweight secondary studies like RRs holds only in scenarios where time and costs are hard constraints. This kind of scenario is typically observed in the practice of many fields. Therefore, RRs are only conceived bounded to practical problems, and conducted within their practical contexts. Thus, practitioners should be willing to devote part of their busy schedule in order to participate on RRs, although the level of participation can vary. RRs that are either conducted without practitioners' collaboration nor related to a problem that emerged from a practical context are considered deviations, and then, should be avoided by the software engineering community.
\end{shaded}

\vspace{0.2cm}
\begin{shaded}
\MyPara{Rapid Reviews are intend to reduce costs and time of heavyweight methods:} To better fit in the practitioners' agenda, RRs should be conducted and reported in a timely manner. Many strategies have been applied to RRs in health-care related fields to reduce costs and time, such as: limiting search strategy by date of publication and/or search source; using just one person to screen studies; not conducting quality appraisal of primary studies;  presenting results with no formal synthesis,among others~\citep{tricco2016international,Tricco2015}.
\end{shaded}

\vspace{0.2cm}
\begin{shaded}
\MyPara{Rapid Reviews results should be reported through mediums appealing to practitioners:} One important aspect of RRs is the way they are reported. Many authors argue that alternative mediums should be used --- when practitioners are the target audience --- instead of the traditional research paper format~\citep{Beecham2014,Grigoleit2015,Cartaxo2016briefings}. To substantiate this claim, \cite{Tricco2015} observed that, although RRs present several variations on their methods and terminologies, 78\% present results as a narrative summary reported in mediums that better fit practitioners' needs. Examples of alternative mediums include: the Contextual Summaries of \cite{Young2014}, that limits the report to a one-page document; the Briefings presented by \cite{Chambers2012}, that summarize the main findings of a secondary study in one section; or even the Evidence Summaries by \cite{Khangura2012}, which use an informative box separated from the main text to highlight the audience and nature of the report. In the context of software engineering, there are few approaches that can be used in this regard. We particularly advocate in favor of the Evidence Briefings (Sect.~\ref{subsec:reporting_briefings}) as a potential way to report the results of a RR.

\end{shaded}

It is important to note that RRs are neither (1) ad-hoc literature reviews, nor (2) an excuse for absence of scientific rigour. RRs must be systematic, by means of following a well-defined protocol. In addition, all the methodological concessions made to a RR must be documented in its protocol. On the RRs report, there must also be a disclaimer about potential methodological limitations (although the details can go on the protocol only, aiming to make the report as concise as possible).

\subsection{Why one should use Rapid Reviews?}

The emerging character of RRs can be explained in terms of its benefits. For instance, a study observed that RRs saved approximately \$ 3 millions when implemented in a hospital~\citep{Mcgregor2005}. Moreover, a survey exploring the use of 15 RRs revealed that 67\% were used as reference material and 53\% were used to, in fact, support decision-making in practice~\citep{hailey2009}. Additionally, \cite{lawani2017five} reported that RRs enabled the development of clinical tools more rapidly than with SRs. Other studies have also demonstrated positive impact of RRs in practice~\citep{Taylor-Phillips2018, hailey2000,batten2012,zechmeister2012,Tricco2015}. Although the main targets of RRs are practitioners, some benefits to researchers and the research community as a whole can be identified. For example, RRs can support and facilitate applied research, or serve as platform to make software engineering research more relevant~\citep{Beecham2014}.

\subsection{Who is using Rapid Reviews?}

Although RRs are not well-known in the software engineering, there is a growing interest in RRs in health-related fields. For instance, \cite{Tricco2015} mapped 100 RRs published between 1997 and 2013 in medicine. Additionally, major medicine venues, such as the prestigious Systematic Reviews journal\footnote{\url{https://systematicreviewsjournal.biomedcentral.com}} officially recognized RRs as one of the Evidence-Based Practice methods~\citep{Moher2015}. Moreover, Cochrane --- a global renowned group of researchers and practitioners specialized in evidence diffusion in health-care --- announced in 2016 a group to play a leading role in guiding the production of RRs~\citep{garritty2016cochrane,CochcraneRapidReviewsGroup}. Due to the increasing importance of RRs, the Canadian Agency for Drugs and Technologies in Health (CADTH), promoted the Rapid Review Summit in 2015, which was focused on the evolving role and practices of RRs to support informed health care policy and clinical decision-making~\citep{polisena2015rapid}. Even the World Health Organization (WHO) has recently published a guide presenting the importance of RRs~\citep{tricco2017rapid}.

\subsection{How Rapid Reviews are compared to Systematic Reviews?}

Some studies were conducted to evaluate the impact of the RRs methodological adaptations by comparing them with SRs. A scoping review found nine studies comparing the results of RRs and SRs. The conclusion shows that their results are both generally similar~\citep{abou2016methods}. To illustrate, \cite{corabian2002rapid} compared six RRs with their SRs peer reviewed publications. The conclusions differed only in one case. Another example is the study of \cite{Best1997}, where two of the RRs they conducted were in agreement with SRs published later on the same topic. Still, \cite{Taylor-Phillips2018} conducted a RR and a SR about the same topic in order to compare their results. The comparison shows that RRs can provide similar results compared to SRs. In that case, both RR and SR identified the same set of papers.

Although there is evidence reporting the similarity of results obtained with RRs and SRs, there is also evidence on the opposite side. For instance, the work of \cite{Van2011} compared results from their RR to a SRs that was conducted by another group, on the same topic, and conflicting results were observed. Therefore, further investigations are still needed to draw more conclusive results.

\begin{shaded}
\MyPara{Rapid Reviews should not be considered as replacements for Systematic Reviews:} We believe RRs should be understood as a complementary scientific product. More concretely, while SRs are important to curate in-depth knowledge, RRs are important to easily and quickly transfer established knowledge to practice.
\end{shaded}

Table~\ref{tab:rrs_vs_SRs} compares the main methodological characteristics of RRs and SRs. The RRs characteristics are based on many medicine studies and guidelines~\citep{tricco2017rapid,Khangura2012,abou2016methods,Taylor-Phillips2018}, while the SRs characteristics are based on Kitchenham's software engineering guidelines~\citep{Kitchenham2007,Cruzes2011,Santos2013}.

\begin{table}[!ht]
    \centering
    \footnotesize
    \caption{Comparison of Rapid Reviews with Systematic Reviews methodological characteristics.}
    \label{tab:rrs_vs_SRs}
    \begin{tabular}{l | l | l}

    \hline
    \multicolumn{1}{c|}{\textbf{CHARACTERISTIC}} & \multicolumn{1}{c|}{\textbf{RAPID REVIEWS}} & \multicolumn{1}{c}{\textbf{SYSTEMATIC REVIEWS}}\\ \hline\hline

    Problem & \parbox{0.4\textwidth}{Bounded to a practical problem, and conducted within a practical context.} & \parbox{0.4\textwidth}{Can emerge from academic and practical contexts~\citep{Kitchenham2007}. However, SRs focusing on problems emerged from practice are the exception~\citep{Santos2013}.} \\ \hline
    Research Questions & \parbox{0.4\textwidth}{Lead to answers that helps solving or at least attenuating the practitioners problem. Exploratory questions aiming to identify which are the strategies and their effectiveness to deal with practitioners problem are one of the gold standards.} & \parbox{0.4\textwidth}{SRs admit questions aiming to support practitioners decision-making, but also studies that are primarily of interest to researchers, with no practice oriented questions~\citep{Kitchenham2007}.} \\ \hline
    Protocol & \parbox{0.4\textwidth}{Must have a document formalizing the protocol.} & \parbox{0.4\textwidth}{Must have a document formalizing the protocol.} \\ \hline
    Stakeholders Roles & \parbox{0.4\textwidth}{Conducted in close collaboration with practitioners, sometimes even having practitioners responsible for executing some of the steps.} & \parbox{0.4\textwidth}{Despite practitioners participation is possible, researchers usually conduct the entire process.} \\ \hline
    Time Frame & \parbox{0.4\textwidth}{Days or Weeks} & \parbox{0.4\textwidth}{Months or Years} \\ \hline
    Search Strategy & \parbox{0.4\textwidth}{- May use few or just one search source (e.g., Scopus).\\- May limit search by publication year, language, and study design.} & \parbox{0.4\textwidth}{- Multiple sources to search for primary studies are recommended.\\- May also limit search by publication year, language, and study design, although more comprehensive search is recommended.} \\ \hline
    Selection Procedure & \parbox{0.4\textwidth}{- Can be conducted by a single person.\\
    - The inclusions/exclusion criteria can be more restrictive aiming to focus on primary studies conducted in contexts similar to the one motivating the RR. (e.g., studies with small/medium/large companies, with companies in countries under specific laws, with open source projects only, etc)~\citep{tricco2017rapid}} & \parbox{0.4\textwidth}{- Must be conducted in pairs to avoid selection bias.\\
    - Usually is less restrictive regarding specificities of primary studies context, specially when it is a mapping study, broader in scope.} \\ \hline
    Quality Appraisal & \parbox{0.4\textwidth}{Conducted by a single person, or not conducted at all~\citep{tricco2017rapid}.} & \parbox{0.4\textwidth}{Conducted in pairs to avoid threats to validity due to low primary studies quality.} \\ \hline
    Extraction Procedure & \parbox{0.4\textwidth}{Usually conducted by a single person to reduce time and effort.} & \parbox{0.4\textwidth}{Conducted in pairs to avoid extraction bias.} \\ \hline
    Synthesis Procedure & \parbox{0.4\textwidth}{Narrative summaries are the most common way to synthesize evidence~\citep{Tricco2015}.} & \parbox{0.4\textwidth}{More systematic methods should be applied (e.g., meta-analysis, meta-ethnography, thematic analysis, etc), although it is not always the case~\citep{Cruzes2011}.} \\ \hline
    Report & \parbox{0.4\textwidth}{Alternative mediums that better fit practitioners needs (e.g., Evidence Briefings).} & \parbox{0.4\textwidth}{Traditional research paper format.} \\ \hline

\end{tabular}

\end{table}

\section{Examples of Rapid Reviews}
\label{sec:examples}

In this section, we describe two RRs we conducted, so one interested in conducting such kind of study can get acquainted with that research approach. The real problems that these RRs were intended to provide solutions to are related to (1) improve customer collaboration and to (2) improve team motivation. We will use these two RRs as example throughout this chapter.

\subsection{Improving Customer Collaboration}
\label{subsec:examples_teammotivation}

This RR was conducted in collaboration with an innovation institute. At first, we had an interview with the institute's representatives to identify the problems they were facing. Among various software projects, we focused on the one that was having difficulties related to low customer collaboration. The complete and detailed results of this experience is reported in~\cite{Cartaxo2018ease}.

This particular software project was late, and the software team needed either the approval and information from its customers to conclude many of the pending tasks. However, the team was having a hard time to establish a proper communication with their client. To illustrate, one of the participants affirmed that \textit{``emails requesting clarification about requirements take one or two weeks for customer to reply.''}

In this context, we decided to conduct a RR together with the practitioners to provide evidence about strategies that would help them to deal with low customer collaboration. More concretely, each aspect of the RR protocol was discussed with the practitioners (e.g., the research questions, the inclusion/exclusion criteria, etc). Online channels such as Skype and e-mail were frequently used during this step. After selecting 17 primary studies, we summarized the findings in an Evidence Briefing document~\citep{Cartaxo2016briefings}. We also ran a workshop to discuss the findings and to answer additional questions. A full time researcher (experienced in conducting secondary studies) was assigned to conduct this RR, which lasted six days. That time frame comprehends the first interview with the institute representatives to identify their problem, up to the workshop in the end to present and discuss the RR results.

After the workshop, we interviewed practitioners to assess their perception regarding the RR we conducted together with them. Practitioners reported many benefits regarding the use of RRs, such as: the novelty of the approach, the applicability to their problem, the reliability of the content, among other. They also reported that the RR fostered the learning of new concepts. As a shortcoming, however, they found that some findings were not clear in the printed version of the Evidence Briefing --- although they became clearer after discussing with researchers during the workshop~\citep{Cartaxo2018ease}.

We also did a follow up the practitioners two months after the workshop to assess whether they applied some of the strategies and findings reported in the RR. Interestingly, we discovered that practitioners indeed adopted some of the strategies in their daily work habits to improve customer collaboration, such as \textit{Story Owner}, \textit{Change Priority}, and \textit{Risk Assessment Up Front}~\citep{Cartaxo2018ease}.

\subsection{Improving Team Motivation}
\label{subsec:examples_contracting_saas}

This RR was performed in collaboration with a software company that develops educational software products in Recife, Brazil. We first contacted the IT director, who is responsible for all the technological aspects of the company. After presenting the goal of this research, a project manager joined us and discussed problems regarding low team motivation he was facing in one of their projects. Similar to the RR on low customer collaboration, this RR was conducted in close collaboration with the practitioners from the software company (e.g., research questions, protocol, etc). The complete and detailed results of this experience is reported in \citep{Cartaxo2018PhdThesis}.

Thirty five studies were selected and their evidence summarized and reported in an Evidence Briefing document. The results were also presented in a workshop. This RR took eight days of a full time researcher experienced in conducting secondary studies.

When interviewing the practitioners after the workshop, they reported many benefits regarding the use of RRs, such as improvements in team confidence and the reliability on RRs findings. They also demonstrated to be willing to embrace RRs in their own process. This particular finding revealed that practitioners are willing to take the risks of using less rigorous methods, such as RRs, in exchange for evidence delivered in short time frames.

\section{The Rapid Review Process}
\label{sec:process}

\begin{figure}[!ht]
\centering
\includegraphics[width=\textwidth]{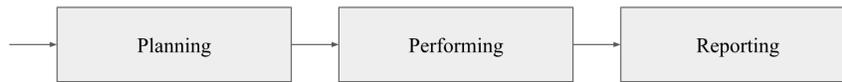}
\caption{Main phases of a Rapid Review.}
\label{fig:process}
\end{figure}

Conducting a RR involves three main phases, as depicted in Fig. ~\ref{fig:process}: planning, performing, and reporting. We describe them in details next.

These phases are similar to the ones of a SR, as described by \cite{Kitchenham2007}. Each phase comprises various specific steps, and that is where the differences between RRs and SRs become evident. While the latter adopts strategies aiming to reduce any type of research bias and to guarantee evidence quality, the former aims to deliver scientific evidence in a timely manner to support practitioners decision-making.

\subsection{Planning a Rapid Review}
\label{sec:planning}

The planning phase of a RR comprehends the creation of a protocol to define all the decisions and procedures demanded to conduct the RR. The protocol must also make explicit the practical problem it intends to provide evidence for, as well as the roles of each stakeholder aiming to guarantee practitioners active participation.

\subsubsection{Demand for a Rapid Review}
\label{subsec:planning_commissioning}

The demand for a RR can emerge from different sources under different contexts. Some possible arrangements we envision are:

\begin{itemize}
    \item \textbf{Practitioners ask for a Rapid Review:} A decision-maker (i.e. practitioner) contacts a researcher or research institution asking for a RR aiming to make decisions based on evidence.

    \item \textbf{Researcher aligns her/his research agenda based on a practical problem:} A researcher contacts a software company (or an open source team) facing problems related to her/his research agenda. A researcher then proposes a RR to both, provide evidence that practitioners need, and to bound her/his research on a practical problem.

    \item \textbf{Researcher prospects a research agenda based on a practical problem:} A researcher contacts a software company (or an open source team) aiming to prospect practical problems to focus her/his research on. In this case, the RR has initially no predetermined focus. To narrow it down, the researcher could leverage interviews with practitioners to grasp the problems they are facing, and then decide which one to attack. This is how we conducted the two RRs presented in Sect.~\ref{sec:examples}.
\end{itemize}

\subsubsection{Defining the Problem}
\label{subsec:planning_problem}

Close collaboration with practitioners is crucial to define the problem that will drive a RR. Since sometimes the problem is not already well-defined (or perhaps not even the practitioner is fully aware of the main problem s/he is facing), researchers can use qualitative research methods such as interviews or focus group to better understand the context and the (eventually hidden) problems~\citep{Cartaxo2018ease}. Depending on how clear is the problem in the practitioners' mind, the interview could be more exploratory (e.g., to understand the whole challenges and needs), more objective (e.g., to understand missing details), or even skipped (e.g., if the problem is very well-defined). One important point to bear in mind when interviewing practitioners to define RRs' problem is that this may be an interactive process. Sometimes you identify a practical problem but there are no studies approaching such problem, so a RR will not be viable, and you may need to find another problem.

\subsubsection{Defining the Research Questions}
\label{subsec:performing_researchquestions}

Research questions in RRs are as important as in SRs~\citep{Kitchenham2007}. Once they are defined, all effort is towards answering them. However, to have useful answers, one has to ask meaningful questions. In RRs, answers are considered useful when they help practitioners to solve or at least attenuate their practical problem. Consequently, questions are considered meaningful only when they lead to such answers.

\begin{shaded}
\MyPara{Research questions in Rapid Reviews should be defined in close collaboration with practitioners:} Questions aiming to identify research gaps or to provide more general insights to the research community should be avoided, left to SRs. RRs should provide answers bounded to the practical context they are inserted into. In other words, they naturally have a narrower character.
\end{shaded}

Each problem will certainly demand different kinds of questions and approaches to investigate them. However, in our experience, exploratory questions aiming to identify strategies to deal with a particular problem are the cornerstone of RRs~\citep{Cartaxo2018ease} since the most important thing to practitioners under time constraints is to discovery strategies, supported by evidence, to solve their problems~\citep{Yourdon1995}.
Examples of such questions are found on the RRs shown in Sect.~\ref{sec:examples}. In the RR about Customer Collaboration we asked:

\begin{itemize}
    \item What are the strategies to improve customer collaboration in software development practice?
    \item What are their effectiveness?
\end{itemize}

Similarly, in the RR about Team Motivation we asked:

\begin{itemize}
    \item What are the strategies to improve software development teams motivation?
    \item What are their effectiveness?
\end{itemize}

Other research questions are possible, if answering them helps practitioners towards the solution of their problem. For instance, in the RR about Customer Collaboration, we also added the following two research questions:

\begin{itemize}
    \item What are the benefits of customer collaboration in software development practice?
    \item What are the problems caused by low customer collaboration in the software development practice?
\end{itemize}

Answers to those questions are useful because the findings were used by the development team to convince their customers about the importance of a better collaboration. On the other hand, these research questions were not necessary on the RR about Team Motivation, since the problem was internal to the company, and the stakeholders already agreed with the importance to improve team motivation. They just did not know how they can do it effectively.

\subsubsection{Defining the Stakeholders Roles}
\label{subsec:planning_stakeholders}

A RR is a joint initiative between researchers and practitioners. Thus, active participation of both sides are not only important, but (as we see it) mandatory. The \textbf{researchers role} is to guarantee the methodological consistency and transparency, while the \textbf{practitioners role} is to make sure that the research is bounded to an actual practical problem, so the evidence will be useful.

In that context, different levels of participation are possible. Considering the extremes, it is possible for researchers to perform all activities related to the RR (e.g., defining the protocol, selecting primary studies, extracting data, synthesizing evidence, and reporting the results) as long as practitioners are involved in the entire process, validating each decision and ensuring the RR is bounded to their practical problem. We could also perceive, nevertheless, that practitioners could perform all RR's activities, as long as researchers are involved, in particular, validating each methodological decision. Any level of participation between these two extremes are also possible and encouraged. However, the effort of each stakeholder will be defined taking into account the time constraints and resources limitations in each specific situation.

Both, the RRs about Customer Collaboration and Team Motivation were conducted near the extreme where researchers defined and executed the reviews. However, the practitioners were aware of every single step made, validating and making suggestions to it. This alignment between researchers and practitioners is crucial to researchers (who conduct the review) do not lose focus, which could lead to, say, research questions that, although interesting from an academic perspective, are not related to a practical problem.

Since RRs and even SRs are not well-known in practice~\citep{Cartaxo2017msr}, we believe this kind of arrangement (where researchers perform most of the RRs tasks) will happen more frequently, at least in the beginning. However, if the collective effort to make software engineering research closer to practice unfolds, then we believe practitioners will recognize the relevance of initiatives like RRs and will be more willing to actively participate.

\subsubsection{Creating the Protocol}
\label{subsec:planning_protocol}

The protocol of a RR has the same goal as the protocol of a SR: to specify all the methodological steps that undertake the review. The protocol itself is one of the most important elements that makes both RRs and SRs systematic. In this sense, it is important to highlight that RRs are not synonymous of ad-hoc literature reviews, but rather systematic. As a consequence, a RR demands a well-documented protocol.

A major difference between RRs and SRs protocols, nevertheless, is the natural inclination of the former to suffer changes throughout the review process. These changes might happen due to the flexible process that RRs allow. However, changes made after the protocol definition must be documented and justified transparently~\citep{tricco2017rapid}.

The components of a RR protocol are similar to the ones of SRs as described by \cite{Kitchenham2007}, such as: research questions, search strategy, inclusion/exclusion criteria, selection procedure, extraction procedure, synthesis procedure, reporting, among others.

Again, we want to highlight the importance of establish a close collaboration with practitioners when defining and conducting a RR protocol. This is crucial to make sure practitioners needs are well-covered and the RR will be performed aiming to provide useful answers. An example of a RR protocol can be found in~\cite{Cartaxo2018ease}.

\subsection{Performing a Rapid Review}
\label{sec:performing}

In this section we present some strategies that may be used to reduce time and cost of performing a RR. For each step, we present some suggestions on how to perform the step. However, one does not have to embrace all strategies, on the contrary, the researcher has to analyze the context and limitations where a RR is being conducted and define which strategies better conciliate given trade-offs. For instance, a RR may use more than one search sources to identify primary studies if ensuring wide coverage is critical, but skip the quality appraisal. While other RR may use just one search source and conduct a rigorous quality appraisal, if the reliability on the evidence is critical.

\begin{shaded}
\MyPara{Transparency is the golden standard in Rapid Reviews:} Regardless the strategies employed to reduce cost and/or time to conduct a RR, limitations and threats to validity must be reported on the protocol. Practitioners may and are willing to consume evidence based on less rigorous methods like RRs, as long as they are aware of the limitations and threats to validity~\citep{Cartaxo2018ease}.
\end{shaded}

\subsubsection{Search Strategy}
\label{subsec:performing_searchstrategy}

SRs usually employ multiple search strategies to guarantee exhaustive coverage such as, using multiple search engines, manual search on conference proceedings and journal issues, as well as forward and backward snowballing approaches.

Adopting all these strategies simultaneously can be extremely resource consuming. RR, on the other hand, may choose to focus on a single search strategy. For instance, instead of using several search engines, RRs may focus on a single one, more likely Scopus or Google Scholar. These search engines cover a wide spectrum of research papers, and usually index papers from the major digital libraries. Complementing the results of the search engine with a snowballing approach has also shown to be a viable option~\citep{Badampudi2015}.
There are other approaches that, if employed, could reduce the effort placed on conducting RRs, such as:

\begin{enumerate}
    \item Limiting the search by date;
    \item Restricting the language in which the paper is written;
    \item Focusing on a given geographical area, or;
    \item Limiting the primary studies according to their research method (e.g.,controlled experiments only, or case studies only)~\citep{tricco2017rapid}.
\end{enumerate}

It is important to note that these approaches may lead to relevant studies being not included, then reducing RR's potential coverage. If one of these strategies are adopted, threats to validity must be transparently reported.
In both the RRs about Customer Collaboration and Team motivation, we used one search source only: the Scopus search engine.

\subsubsection{Selection Procedure}
\label{subsec:performing_selectionprocedure}

Since RRs are bounded to a practical context, one may define restrictive inclusion/exclusion criteria. The goal here is twofold: to reduce the amount of studies to screen and to provide evidence that better fit practitioners needs.

For instance, the RR about Team Motivation was conducted in a small private company with collocated teams. Therefore, some of the inclusion/exclusion criteria were the following:

{\footnotesize
    \begin{itemize}
        \setlength{\itemindent}{+.25in}
        \item The study must not be related to large companies;
        \item The study must not be related to distributed teams;
        \item The study must not be related crowd source software development;
        \item The study must not be related to open source software development;
    \end{itemize}
}

Defining restrictive inclusion/exclusion criteria may reduce the time and effort to conduct a RR. However, this procedure does not necessarily incur in threats to validity. In fact, this may be considered a good practice, when the restrictions are made aiming to provide evidence only from primary studies conducted in contexts similar to the one the RR is being conducted. Highly contextualized studies are long considered one of the best ways to have impact in practice~\citep{Dyba2012,Cartaxo2015}.

Moreover, SRs usually require independent screening of studies by at least two reviewers~\citep{Kitchenham2007,tricco2017rapid}, which is very resource intensive. RRs, on the other hand, may have a selection procedure conducted by a single reviewer. Another option is to have a second reviewer just to pass through a reduced sample of studies. Such strategies may obviously introduce selection bias and must be reported accordingly.

Usually, SRs splits the selection procedure in two substeps. In the first, reviewers screen primary studies' titles and abstracts, and in the second, the entire papers content. To abbreviate this process, one may split the selection procedure in three substeps, instead of two. The first substep can be dedicated to screening primary studies' titles only. This might accelerate the exclusion of papers that are clearly out of scope since it prevent one to read papers abstracts. On the other side, it may provoke false negatives. The second substep would select primary studies based on abstract only, and the third sub-step based on the entire content. Regarding this particular strategy, one of the practitioners that participated on the RR about Customer Collaboration give us the following feedback:

\begin{quote}
    ``Sometimes we search for solutions in just one source [...] Then we do it exactly as recommended by that source but it may not work for us. When we do it like this [the RR], we can have more possibilities [the strategies identified by the RR], even considering it was conducted faster [the RR compared to SRs], and maybe many things [papers] could be lost just because of the title [the first round of selection procedure, which we analyzed only the titles of the papers], because someone put a bad title. That is ok, who cares?''
\end{quote}

\subsubsection{Quality Appraisal}
\label{subsec:performing_qualitycriteria}

In addition to inclusion/exclusion criteria, quality criteria are also usually defined in SRs in order to select high quality evidence only. In a more extreme view, RR researchers can entirely skip this step, but threats to validity associated with this decision must be transparently reported. Both the RRs we presented in Sect. \ref{sec:examples}, adopted this strategy.

Another less radical strategy would be to focus only on studies published on conferences and/or journals that employ a rigorous review process. This may increase the chances of selecting high quality evidence with a low effort (e.g., no need to analyse the evidence quality of each and all papers). Although this approach can also have limitations (e.g., a potentially relevant study could have published on a less prestigious venue or on arXiv), at least we know that the primary studies being included already passed through a rigorous sieve.

If evidence quality is critical in the context where the RR is being conducted, a strategy that may reduce the time and effort is to have quality appraisal carried out by a single reviewer or using pairs to appraise just a sample of papers. In contrary to SRs, where quality appraisal is recommended to be conducted fully in pairs.

\subsubsection{Extraction Procedure}
\label{subsec:performing_extactionprocedure}

The data extraction procedure can be conducted by a single reviewer in RRs, as long as the inherent biases are transparently reported. Both the RRs we presented in Sect. \ref{sec:examples}, adopted this strategy. Moreover, in SRs, when data is missing on the selected studies, it is usually recommended to contact the authors. Researchers who conducted RRs in medicine very infrequently indeed contacted primary studies' authors~\citep{tricco2017rapid}. That can be a viable strategy: studies with missing data should probably be excluded from the RR, and their exclusion must be reported. RRs consumers (a.k.a practitioners) can reach those studies later if they wish to.

\subsubsection{Synthesis Procedure}
\label{subsec:performing_synthesisprocedure}

Knowledge synthesis is probably one of the most important steps of any secondary study, but at the same time one of the most time consuming activities. However, a tertiary study revealed that as many as half of the SRs analyzed in software engineering do not present any kind of formal knowledge synthesis procedure~\citep{cruzes2011research}. They also summarized various methods for knowledge synthesis (e.g.,meta-analysis, meta-ethnography, grounded theory, qualitative metasummary, among others) to encourage researchers to apply them.

A possible strategy to reduce time and effort synthesising evidence in RRs is using lightweight methods, like Narrative Synthesis~\citep{cruzes2011research,tricco2017rapid}, in contrast to the more rigorous and time/effort consuming ones, like Meta-Analysis~\citep{lipsey2001practical} or Grounded Theory~\citep{Stol2016} methods alike. This decision brings an obvious limitation and must be reported, so practitioners consuming RRs evidence can make informed decision.

Conclusions, recommendations, and implications are particularly important in RRs since they can guide practitioners to adopt the synthesized knowledge. In medicine, they encourage researchers to dedicate time to make her/his conclusions and recommendations to practitioners, and avoid presenting a report with findings only~\citep{tricco2017rapid}. We experienced such kind of demand from practitioners on the RR about Team Motivation, when a practitioner gave us the following feedback:

\begin{quote}
    ``since it [the RR] was focused on our problem, maybe if there was something saying which one [strategy identified with the RR] you recommend [...] this is what is missing [...] maybe it is missing a conclusion, the researcher's comments.''
\end{quote}

In addition, one should keep in mind that those conclusions, recommendations, and implications should be strongly bounded to RRs context, in opposition to the ones draw with SRs that usually aims to reach a wider audience and scope~\citep{tricco2017rapid}.

\subsection{Reporting a Rapid Review}
\label{sec:reporting}

Reporting and disseminating knowledge produced with RRs are as important as conducting the RR itself. SRs are usually conducted in academic environment and thus the report is usually focused on that audience. That means SRs are commonly reported in scientific paper format and diffused through academic journals and conferences.

RRs, however, target software practitioners. Therefore, one should consider that not all information that is crucial to researchers is also relevant to practitioners (e.g., research method, background, related work, etc). As a consequence, RRs must be reported in a more straightforward way, focusing on results and recommendations, so practitioners can easily consume the information to support their decision making.

There are several approaches that could be used in this regard, as presented in Sect. \ref{subsec:what_is_rr}~\citep{Chambers2012,Khangura2012,Young2014,Best1997}.
This section presents the concept of Evidence Briefings, which are alternative mediums to report RRs more focused on practitioners needs, and also discusses the importance of disseminating knowledge produced with RRs.

\subsubsection{Evidence Briefings}
\label{subsec:reporting_briefings}

Evidence Briefings are one-page documents reporting the main findings of RRs~\citep{Cartaxo2016briefings}. A template, as well as examples of such documents can be found online\footnote{\url{http://cin.ufpe.br/eseg/briefings}}. The Evidence Briefings template was defined based on the best practices observed in medicine as well as on Information Design~\citep{Tondreau2011} and Gestalt Theory~\citep{Lupton2015} principles. Figure~\ref{fig:briefings_structure} shows an example of an Evidence Briefing. The numbers within squares denote each part of Evidence Briefing's structure, and following there are some guidelines on how to fill each of those parts:

\begin{figure}[!ht]
    \centering
    \includegraphics[width=0.5\linewidth]{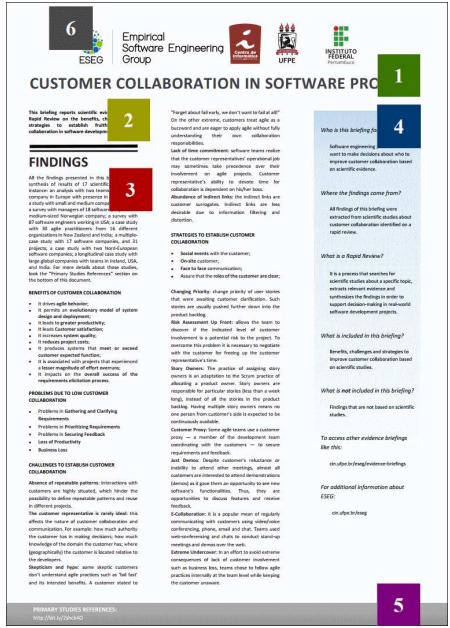}
    \caption{\noindent Evidence Briefing structure.}
    \label{fig:briefings_structure}
\end{figure}

\begin{enumerate}
\item The \textbf{title} of an Evidence Briefing should be as concise as possible. Usually, one or two lines titles. Titles with more than two lines should be avoided since they might reduce document space to report RRs' findings.

\item To fill the Evidence Briefing's \textbf{summary}, we suggest researchers to adopt the following structure: \textit{This briefing reports scientific evidence on $<$RESEARCH GOAL$>$}. The summary should span few lines. Following is an example of Evidence Briefing's summary: \textit{``This briefing reports scientific evidence on the challenges involved in using Scrum for global software development (GSD) projects, and strategies available to deal with them.''}

\item The \textbf{findings} section is the most important one. It should list the main findings of the RR. When writing the findings, we recommend to use one finding per paragraph. Bullets to highlight important points as well as charts, figures, and tables are welcome since they make the findings even easier to read. Findings should be short sentences, straight to the point. The findings section should not have information about the research method. The idea of the Evidence Briefing is to quickly communicate the main findings of a RR to practitioners. If they have interest they can refer to the complementary material reference shown in the item 5.

\item The \textbf{box at the right side} of the Evidence Briefing should be filled with information about the Evidence Briefing's target audience, clarifications about what information is included, and what is not included in the Evidence Briefing. The template has a complete set of suggestions to structure information at the right box.

\item The \textbf{reference} to complementary material should be placed at the bottom of the Evidence Briefing. It may be a link to a webpage containing at least the following documents/information: the RR protocol document and a list of references to the primary studies included in the RR.

\item \textbf{Logos} of universities, software companies, and any other institutions involved in the RR initiative should be placed at the very top of the Evidence Briefing document. This publicizes the institutions producing Evidence Briefings, and might make practitioners search for more RRs in the institutions websites.
\end{enumerate}

Although other mediums to transfer scientific evidence exist, we recommend the use of the Evidence Briefings because, as observed in an empirical evaluation, both researchers and practitioners are positive about using Evidence Briefings as medium  to transfer scientific knowledge to software engineering practice~\citep{Cartaxo2016briefings}.

\subsubsection{Dissemination of Rapid Reviews Results}
\label{subsec:further_discusions_diffusion}

Not all RRs are disseminated beyond the practitioners scope due to sensitive information belonging to the software company involved. However, if this is not the case, we recommend RR researchers to post the RRs report (e.g., Evidence Briefing) online on the research institution's or the company's website. Sharing the report on social networks such as Twitter or ResearchGate can also increase the impact of the reviews.

\section{Further Discussions on the Feasibility of Rapid Reviews}
\label{sec:further_discussions}

In this section we present further discussions about topics that may concern software engineering research community about the feasibility of RRs as an Evidence-Based method.

\subsection{Research Community Viewpoints on Rapid Reviews}
\label{subsec:further_discusions_viewpoints}

Although RRs are a rising research method in the medical domain, they are so far barely known in the SE community. We believe our community could and ought to benefit from it. However, due to the lack of RRs studies in software engineering, little is known about how our research community perceives the adoption of RRs.

This is particularly important because, according to \cite{rogers2003diffusion}, the perceptions of all individuals involved in an initiative is one of the main predictors of its adoption. The importance of exploring the perceptions of practitioners -- as we have done in~\cite{Cartaxo2018ease} -- is easy to understand since practitioners are the target audience of RRs. But the perceptions of researchers should certainly not be neglected. Moreover, if the software engineering research community discards RRs, such kind of initiative can easily end even before having shown its potential. In informal discussions with EBSE specialists during conferences, we observed that their opinions about RRs seem to be highly polarized, especially when methodological concessions are made.

This feeling is now backed up with evidence provided in a study we conducted with 37 software engineering researchers~\citep{cartaxo2019esem}. We applied a Q-Methodology approach, enabling us to identify that researchers in software engineering can be classified in four groups according to their viewpoints regarding RRs: \\

\MyPara{Unconvinced:} researchers aligned with this viewpoint are the ones that agree the most that further research comparing the methods and results of RRs and SRs is required before they decide how they feel about RRs. The indecision of this viewpoint towards RRs is even more explicit when we look the contradictory affirmations the participants provided. They think a well-conducted RR may produce better evidence than a poorly conducted SRs, but on the other hand, they have more confidence in evidence produced with a SR than in evidence produced with a RR.\\

\MyPara{Enthusiastic:} researchers aligned with this viewpoint are generally favorable about RRs, and believe RRs can provide reasonable evidence to practitioners, if minimum standards to conduct and report RRs are established. They also strongly agree that a well-conducted RR may produce better evidence than a poorly conducted SR.\\

\MyPara{Picky:} researchers aligned with this viewpoint are very skeptical about RRs, as well as concerned about the quality of primary studies included in RRs and how the results are reported. This negative perception can be explained by a strong belief hold by researchers aligned with this viewpoint, which is that knowledge users (practitioners) do not fully understand the implications of RR methodological concessions. Researchers with this viewpoint also put little faith in RRs validity. They strongly disregard the possibility that RRs can be timely and valid, even when methodological concessions are made.\\

\MyPara{Pragmatic} Researchers aligned with this viewpoint pragmatically focus on variety of contextual information to decide if RRs are the best fit to support decision-making. They also believe practitioners are able to understand the impacts of flexible research methods adopted by RRs. Still, they believe rigid standards in RRs could reduce their usefulness to practitioners.\\

Although the viewpoints are quite diverse, there is a consensus around the opinion that both RRs and SRs can be conducted very well or very poorly, and that time needed to conduct an evidence synthesis study is not related to its quality. The main concerns about RRs -- not necessarily shared among the four viewpoints -- are: the need for more evidence about the effectiveness of RRs, the importance to determine minimum standards, the relevance of quality assessment to include primary studies, and the emphasis on transparency in RRs.

With this typology in mind, one can better understand what are the main concerns of researchers and promote better understanding about RRs. As consequence, our community can pave a road better connecting research with practice and make software engineering research more impactful and relevant.

\subsection{Publishing Rapid Reviews in Scientific Peer Reviewed Venues}
\label{subsec:further_discusions_peerreviewed}

Since RRs are commonly reported in non-scientific paper format (i.e. Evidence Briefings), they are usually internally reviewed, but not peer reviewed~\citep{tricco2017rapid}. This may be seem as an unpromising incentive since researchers are paper publishing driven beings. Nevertheless, we encourage researchers who conduct RRs to also publish their results in traditional scientific venues by reporting their results in a scientific format too.

\begin{shaded}
\MyPara{Rapid Reviews can and should also be published in academic peer reviewed venues:} One may argue that a RR will probably not constitute enough contribution to deserve a rigorous scientific publication. However, one should note that RRs are usually inserted into broader knowledge/technology transfer initiatives~\citep{Cartaxo2018ist}, and such initiatives are usually very enriching and welcomed in scientific venues. The paper may report not only the RR protocol and results, but also the perceptions of practitioners participating on the entire RR initiative. One example of such peer reviewed RR publication in software engineering is one of our work~\citep{Cartaxo2018ease}. Additionally, if the cooperation between researchers and practitioners goes beyond the RR itself -- for instance, when researchers actively participate, together with practitioners, designing the solutions to practitioners' problems based on the evidence provided by the RR, and adopting a participatory method like action research -- the paper may report how the knowledge produced with that RR was applied in practice, and in what degree it solved or at least attenuated practitioners problem. In fact, this kind of research would probably close the entire knowledge/technology transfer cycle in a marvelous way. It puts the scientific knowledge in action with direct impact in practice.
\end{shaded}

\subsection{On the use of Grey Literature}

The last point that worth discussing is whether one could conduct a RR with grey literature. The is a positive argument along these lines, which is often related to how practitioners share and acquire knowledge (i.e., through blog posts, talks, videos, etc). These mediums are often created by (and for) practitioners, and do not necessarily pass through a rigorous revision process. Although more recently some researchers are taking advantage of grey literature on their on study \citep{garousi2017guidelines,Garousi2016multivocal}, there is still some conservative researchers that favor the traditional peer reviewed literature. In this chapter, we do not intent to add more fire on this already heated debate. However, we also concur that eventually, a researcher conducting a RR would have to think about what kind of literate s/he will include on her review.  To guide this researcher, our experience suggests that researchers should focus only on peer reviewed literature when conducting a RR. This is particularly due to the fact that RRs have already several limitations and threats to validity. We believe that adding grey literature to this equation could weaken the quality of the review produced, at least in the eyes of an unconvinced researcher. Obviously, this is an hypothesis that could be tested in follow up studies. For more detailed information about using Grey Literature as evidence, refer to Chap. 14.

\section{Recommended Further Reading}
\label{sec:further_reading}

For a better comprehension of this chapter, we suppose the reader have experience conducting SRs, or at least have a good notion of what a SR is, as well as the steps and procedures it comprises. If that is not the case, please refer to \cite{Kitchenham2007} guidelines as well as \cite{Kitchenham2004} EBSE seminal paper.

Regarding RRs, one can read the first experience conducting such kind of study in software engineering in \cite{Cartaxo2018ease}. We also recommend reading the practical guide on RRs provided by the World Heath Organization~\citep{tricco2017rapid}. It distills most of the accumulated experience conducting RRs in medicine. For a comprehensive view on the state of practice and research about RRs in medicine, one can take a look on \cite{Tricco2015} scoping study. It analyzes 100 RRs conducted between 1997 and 2013 under various perspectives, such as RRs characteristics, terminology, citation, impact on practice, comparison with SRs, among others. For a better understanding on how RRs fit in a more comprehensive knowledge/technology transfer initiative, there is our study proposing such a model in \cite{Cartaxo2018ist}.

Regarding initiatives related to RR, there is a recent trend towards the use of gray literature in Multivocal Literature Reviews (MLRs)~\citep{Garousi2016multivocal,garousi2017guidelines,yasin2012quality}. Generally speaking, the use of MLRs shares the core goal of a RR, which is to make research more aligned with practice. However, there is a fundamental difference between these two approaches. On the one hand, RRs aims to provide knowledge based on scientific evidence from peer-reviewed and rigorous primary studies only, as well as deliver evidence in a timely manner. On the other hand, MLRs applies systematic methods to synthesize not only primary studies, but also gray literature. Moreover, MLRs do not necessarily emerge from a practical problem nor is necessarily concerned about delivering evidence in a timely manner to practitioner. Thus, RRs and MLRs are different approaches, although both can potentially contribute to reduce the gap between software engineering research and practice.

\section{Conclusion}
\label{sec:conclusion}

A new era of software engineering has emerged, and it is changing the way we think about empirical research. In a recent series of posts at Communications of ACM blog, Bertrand Meyer~\citep{meyer2018theendofsoftware,meyer2018empiricalanswers1,meyer2018empiricalanswers2} precisely framed this era throughout a vision where empirical evidence and practice orientation are pivotal elements:

\begin{quote}
\textit{``As long as empirical software engineering was a young, fledgling discipline, it made good sense to start with problems that naturally landed themselves to empirical investigation. But now that the field has matured, it may be time to reverse the perspective and start from the consumer's perspective: for practitioners of software engineering, what problems, not yet satisfactorily answered by software engineering theory, could benefit, in the search for answers, from empirical studies?''}~\citep{meyer2018empiricalanswers1}
\end{quote}

Meyer's voice certainly is not alone. Many other researchers are starting to recognize practice orientation as the next long way ahead~\citep{Beecham2014,Duarte2015,Laird2015, Santos2013}. Unfortunately, there is evidence that secondary studies in software engineering lack connection with practice~\citep{Santos2013,daSilva2011,Hassler2014,Cartaxo2017msr}.

In this chapter, we introduce the concept of Rapid Reviews (RRs) in the context of knowledge transfer in software engineering. They are secondary studies aiming to provide research evidence to support decision-making in practice, and in consequence, must be conducted taking into account the constraints inherent to practical environments. RRs usually deliver evidence in a more timely manner, with lower costs, reporting results through more appealing mediums, and more connected to practice, when compared to Full Systematic Reviews.

We also present examples of experiences conducting RRs together with software engineering practitioners. They affirmed to have learned new concepts about the problem they were facing, as well as declared to trust in the findings provided with the RRs.
We also present guidelines covering the entire RRs process aiming to help researchers and/or practitioners interested in conducting their own RRs.

Even looking for all the good results, one to be fair has to highlight that RRs are not always a bed of roses. RRs have their limitations, and this must be considered carefully. They are certainly not silver bullets nor they are substituting Systematic Reviews. Moreover, we explore and provide solutions aiming to address some concerns that researchers may have about the feasibility of RRs as a viable Evidence-Based research method, like: researchers perceptions (scepticism) about RRs flexible strategies, how to publish RRs in scientific rigorous peer review venues, as well as how to disseminate the results obtained with the RRs.

In conclusion, we believe RRs can play an important role on promoting knowledge transfer from scientific empirical evidence to practice, and reduce the gap between academic research and software engineering practice.

\bibliographystyle{agsm}
\bibliography{references}

\end{document}